\documentclass[fleqn,usenatbib]{mnras}

\usepackage{newtxtext,newtxmath}
\usepackage[T1]{fontenc}

\DeclareRobustCommand{\VAN}[3]{#2}
\let\VANthebibliography\thebibliography
\def\thebibliography{\DeclareRobustCommand{\VAN}[3]{##3}\VANthebibliography}

\usepackage{graphicx}	% Including figure files
\usepackage{amsmath}	% Advanced maths commands
\usepackage[utf8]{inputenc}
\usepackage{indentfirst}
\usepackage{hyperref,graphicx} 
\usepackage{mathrsfs} 
\usepackage{amsmath,siunitx}
\usepackage{bm}
\graphicspath{{./fig/}}
\usepackage{natbib}
\usepackage[version=4]{mhchem}
\usepackage[table]{xcolor}

\newcommand{\acri}{a_{\text{critical}}}
\newcommand{\afrag}{a_{\text{frag}}}
\newcommand{\cs}{c_{\mathrm{s}}}
\newcommand{\cso}{c_{\mathrm{s, 0}}}
\newcommand{\dlogpr}{\mathrm{d}\log P/\mathrm{d}\log r}
\newcommand{\ftrap}{f_{\rm{trap}}}
\newcommand{\gw}{\Delta_{\mathrm{gap}}}
\newcommand{\gwn}{\Delta_{\mathrm{gap}}/r_{\mathrm{p}}} % normalized gap width
\newcommand{\gd}{\Sigma_{\rm{gap}}/\Sigma_{0}}
 
\newcommand{\hpr}{H/r}

\newcommand{\Me}{M_\oplus}
\newcommand{\Msun}{M_\odot}
\newcommand{\Mstar}{M_\star}
\newcommand{\Mp}{M_{\rm{p}}}
\newcommand{\Mj}{M_\mathrm{J}}

\newcommand{\nrhd}{N_{\rm{r, HD}}}
\newcommand{\nphihd}{N_{\rm{\phi, HD}}}

\newcommand{\nrrt}{N_{\rm{r, RT}}}
\newcommand{\nphirt}{N_{\rm{\phi, RT}}}
\newcommand{\nthetart}{N_{\rm{\theta, RT}}}
\newcommand{\nphot}{N_{\rm{photon}}}

\newcommand{\rdt}{r_{\rm{dt}}}

\newcommand{\rgapo}{r_{\rm{gap, out}}}
\newcommand{\rhill}{R_{\mathrm{H}}} %hill radius
\newcommand{\rmin}{r_{\rm{min}}}
\newcommand{\rmax}{r_{\rm{max}}}
\newcommand{\rp}{r_{\mathrm{p}}} % planet location
\newcommand{\Rstar}{R_\star}
\newcommand{\Rsun}{R_\odot}

\newcommand{\sigmag}{\Sigma_{\rm{g}}}

\newcommand{\sigmagap}{\Sigma_{\rm{gap}}}
\newcommand{\Tmid}{T_{\rm{mid}}}
\newcommand{\Tstar}{T_\star}

\newcommand{\Tsub}{T_{\rm{sub}}}
\newcommand{\Tgas}{T_{\rm{gas}}}
\newcommand{\Tdust}{T_{\rm{dust}}}
\newcommand{\Twater}{T_{\rm{sub, H_{2}O}}}
\newcommand{\Tcd}{T_{\rm{sub, CO_{2}}}}
\newcommand{\Tcm}{T_{\rm{sub, CO}}}
\newcommand{\Tratio}{\Delta T/T_{\rm{it100}}}
\newcommand{\vf}{v_{f}}

\newcommand{\fargo}{\text{FARGO3D}}
\newcommand{\hd}{\text{hydrodynamical}}
\newcommand{\ite}{\text{iteration}}
\newcommand{\noit}{\text{non-iteration}}
\newcommand{\ppd}{\text{protoplanetary disk}}
\newcommand{\radmc}{\text{RADMC-3D}}
\newcommand{\rt}{\text{radiative transfer}}

\definecolor{lukecolor}{RGB}{143, 53, 53}

\title[Gap-opening feedback]{Planet Gap-opening Feedback on Disk Thermal Structure and Composition}

\author[K. Chen et al.]{
Kan Chen,$^{1}$\thanks{E-mail: kan.chen.21@ucl.ac.uk}
Mihkel Kama,$^{1,2}$
Paola Pinilla$^{3}$
and Luke Keyte$^{1}$
\\
$^{1}$ Department of Physics and Astronomy, University College London, Gower Street, London, WC1E 6BT, UK\\
$^{2}$ Tartu Observatory, University of Tartu, Observatooriumi 1, Tõravere 61602, Tartu maakond, Estonia\\
$^{3}$ Mullard Space Science Laboratory, University College London,
Holmbury St Mary, Dorking, Surrey RH5 6NT, UK
}
\date{Accepted XXX. Received YYY; in original form ZZZ}

\pubyear{2022}

\begin{document}
\label{firstpage}
\pagerange{\pageref{firstpage}--\pageref{lastpage}}
\maketitle

% \input{mycommand.tex}

% Abstract of the paper
\begin{abstract}
(Exo-)planets inherit their budget of chemical elements from a protoplanetary disk. The disk temperature determines the phase of each chemical species, which sets the composition of solids and gas available for planet formation. We investigate how gap structures, which are widely seen by recent disk observations, alter the thermal and chemical structure of a disk. Planet-disk interaction is a leading hypothesis of gap formation and so such changes could present a feedback that planets have on planet-forming material. 
Both the planet gap-opening process and the disk thermal structure are well studied individually, but how the gap-opening process affects disk thermal structure evolution remains an open question. 
We develop a new modelling method by iterating hydrodynamical and radiative transfer simulations to explore the gap-opening feedback on disk thermal structure. We carry out parameter studies by considering different planet locations $\rp$ and planet masses $\Mp$. We find that for the same $\rp$ and $\Mp$, our iteration method predicts a wider and deeper gap than the non-iteration method. We also find that the inner disk and gap temperature from the iteration method can vary strongly from the non-iteration or disk without planets, which can further influence dust-trap conditions, iceline locations, and distribution of various ices, such as \ce{H2O}, \ce{CO2}, and CO on large dust grains (“pebbles”). Through that, a gap-opening planet can complicate the canonical picture of the non-planet disk C/O ratio and influence the composition of the next generation of planetesimals and planets.

\end{abstract}

% Select between one and six entries from the list of approved keywords.
% Don't make up new ones.

%%%%%%%%%%%%%%%%%%%%%%%%%%%%%%%%%%%%%%%%%%%%%%%%%%

%%%%%%%%%%%%%%%%% BODY OF PAPER %%%%%%%%%%%%%%%%%%

\section{Introduction}

Chemical element abundance ratios in planets, and in comets or asteroids, are determined by the chemical composition and physical-chemical evolution of the protoplanetary disk they form in. The study of chemical element abundance ratios such as C/O \citep{oberg_effects_2011, madhusudhan_toward_2014} or N/S \citep{turrini_tracing_2021} may allow to connect planetary bodies to their formation history, which is important for understanding how the chemical diversity of planetary systems arises. 
The distribution of volatile chemical elements in the solid (dust, ice) and gas phases is set by the location of their icelines, which depend on the disk temperature structure. In this work, we employ hydrodynamical and radiative transfer models to study the feedback of planet-induced gaps on the temperature structure and hence the location of icelines.

ALMA observations have revealed that rings and gaps in the dust and gas components are common in protoplanetary disks \citep[e.g.,][]{andrews_disk_2018, oberg_molecules_2021}. 
One possible and intriguing explanation for the formation of such substructures is embedded young planets in disks. Despite great efforts, very few protoplanets have been detected in disks by direct imaging \citep{keppler_discovery_2018, keppler_highly_2019, haffert_two_2019, benisty_circumplanetary_2021, currie_images_2022, hammond_confirmation_2023}. Direct imaging is, however, biased towards super-Jupiter mass protoplanets, whereas most gaps may be due to lower-mass giant planets. Their masses can be inferred from the gap structure or gas kinematics \citep[e.g.,][]{zhang_disk_2018, teague_kinematical_2018}.
Alternative scenarios to explain gaps and rings without planets have also been proposed, such as secular gravitational instabilities \citep{takahashi_two-component_2014}, dust evolution \citep{birnstiel_dust_2015},
zonal flows \citep{flock_gaps_2015}, and icelines \citep{zhang_evidence_2015}.

A gap in the disk implies a reduced optical depth in a radially confined region. This allows shorter wavelength photons to penetrate deeper and heat the disk midplane, as well as the edges of the gap, so gaps potentially affect the disk temperature structure. An opposite, cooling effect may result from fewer photons being scattered by dust towards the midplane. The balance of these effects around a given dust gap can be studied with Monte Carlo radiative transfer (RT) models \citep{broome_iceline_2022}. Previous studies of temperature changes around gaps used analytically prescribed surface density profiles: \citet{cleeves_indirect_2015} explored the spatial distribution of molecular abundances resulting from increased heating due to an accreting protoplanet in a gap, while \citet{broome_iceline_2022} used Monte Carlo radiative transfer to investigate the dust temperature structure around analytical gap profiles in a hydrostatic 1+1D disk model.

The temperature change caused by a gap can also affect the structure of the gap itself. 
Hydrodynamical (HD) simulations of planet-disk interactions and gap-opening processes assuming a locally isothermal equation of state (EoS) provide empirical formulas of gap depth and width \citep{fung_how_2014, kanagawa_mass_2015, kanagawa_mass_2016,zhang_disk_2018, duffell_empirically_2020}.
Recently, \citet{miranda_planetary_2019, miranda_planetdisk_2020} suggest that the assumptions of the equation of state, locally isothermal or adiabatic assumptions, can affect the gap properties by altering the propagation of density waves. Additionally, \citet{zhang_effects_2020} used simulations to show that the cooling timescale can influence the gap profile.

Disk thermodynamics plays an important role in setting the location of different icelines in disks.
An iceline of a specific molecule is the location where the temperature is low enough so that such molecules freeze out from the gas phase onto dust grains. 
Though direct measurements of the location of molecular icelines are rare in observations (e.g., water iceline \citet{van_t_hoff_imaging_2018}, \ce{CO} iceline \citet{zhang_mass_2017, van_t_hoff_robustness_2017}),
icelines can play an important role in planet formation. Across  icelines, the gas composition and ice reservoirs for the planet and planetesimal formation are changed \citep[e.g.,][]{oberg_effects_2011}, and the efficiency of planetesimal formation can increase at the water iceline \citep[e.g.,][]{stevenson_rapid_1988, schoonenberg_planetesimal_2017}.
In addition, dust trapping is closely related to the planet gap-opening process, which in combination with the location of icelines  determine the location of planetesimal formation and their composition. Dust trapping in local pressure maxima is proposed to overcome rapid dust loss due to radial drift by the drag between the gas and the dust in disks \citet{whipple_certain_1972}. For example, \cite{pinilla_ring_2012} demonstrated that the pressure bump outside the gaps opened by planets can trap dust and produced ring-like structures as observed

Conventionally, previous studies on gap modeling or thermal structures in disks only conduct HD or RT simulations, or combine the final results from HD to RT simulations to compare with observations. However,  as planets open gaps in disks, the temperature around gaps could deviate significantly from the temperature adopted for disks without planets. In the meantime, the temperature changes affect the disk gas scale height $H$ and volume density $\rho$ distribution. 

In this paper, we build a new model to investigate the planet gap-opening process and the gap-opening feedback on disk thermal structure. 
Because the temperature controls which species can exist as solid ices, our model allows us to investigate the question: what is the feedback effect of giant planets on the composition of material subsequently accreted by the planets themselves, or by a new generation of forming planetesimals? 

In order to improve previous models, 
we first feed HD simulations with a more physical energy field from RT models. During the planet gap-opening process, we combine the HD and RT simulations together and iterate them.
We implement the new temperature calculated by RT to correct the energy field of HD simulations.

This paper is organized as follows. In Section \ref{sec:method} we describe our modeling method of how we iterate the $\hd$ and $\rt$ simulations to study the gap-opening process. In Section \ref{sec:result}, we present and quantify our modeling results of gap properties, disk temperature structure, and ice distributions. 
Section \ref{sec:discussion} discusses the impact of our results on disk composition, disk substructure observation, and the limits of our models.
Section \ref{sec:conclusion} summarizes the main conclusions of this paper.

\section{Methods}\label{sec:method}

In this section, we describe the codes and setup of our $\hd$ and $\rt$ simulations, as well as the workflow of how we iterate these two simulations to study the temperature structure of a disk with a gap-opening planet.

\begin{figure*}
\centering
\includegraphics[width=\linewidth]{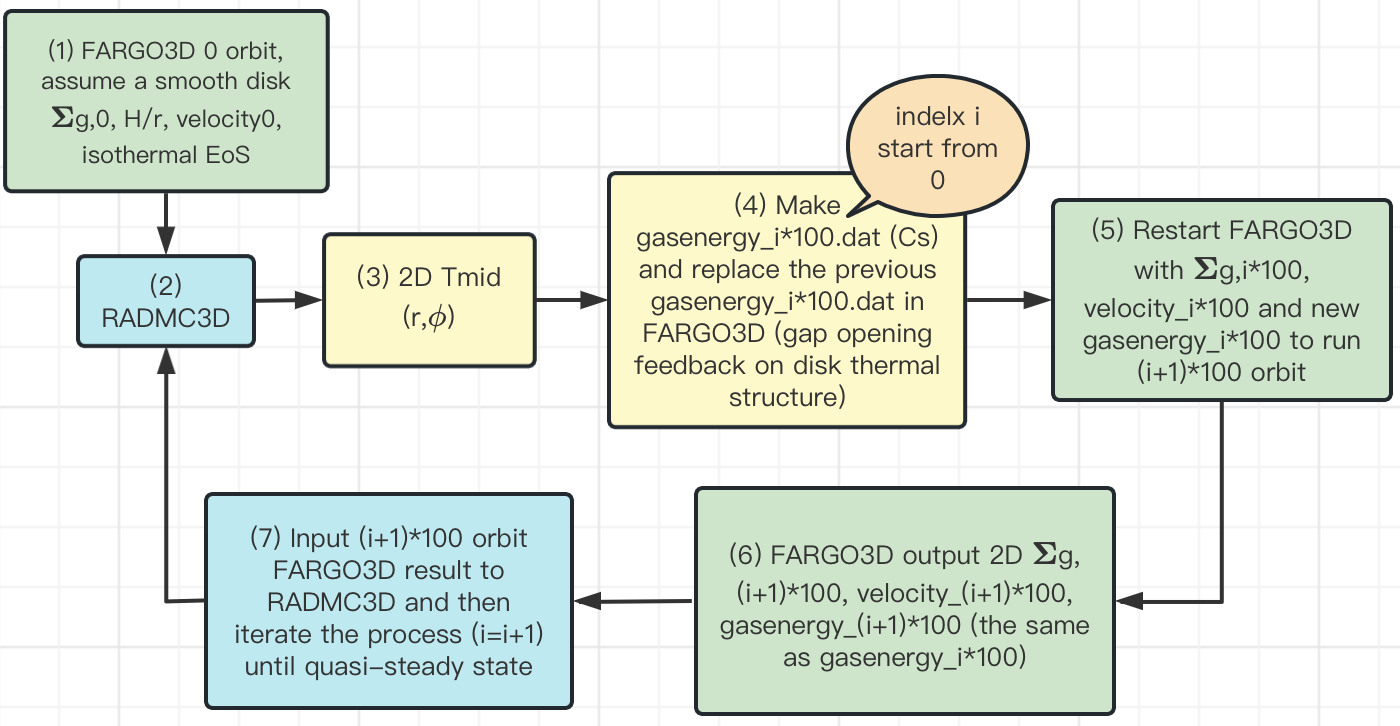} 
\caption{The workflow of the iteration method. The green, blue, and yellow boxes represent the steps of $\fargo$,  $\radmc$, and post-processing between $\radmc$ and $\fargo$, respectively. The iteration step is 100 planet orbits.
}
\label{fig:workflow_it} 
\end{figure*}

%%%%%%%%%%%%%%%%%%%%%%%%%%%%%%%
\subsection{Hydrodynamical simulations}

We conduct 2D $\hd$ simulations in polar coordinate (r, $\phi$) to study surface density evolution with $\fargo$ \citep
{benitez-llambay_fargo3d_2016}. The main parameters of $\fargo$ simulations are shown in Table \ref{tab:fg}.

For grid setup, we conduct global disk simulations of a transition disk which extend from $r_{min} = 1$ au to $r_{max} = 100$ au. 
The global disk simulations for $\fargo$ avoid the radial extrapolation of the sound speed $\cs$ setups for global disk $\radmc$ simulations. Such consideration is necessary, as the extrapolation could be imprecise for a radially non-smooth $\cs$ field.  We set up mesh grids linearly distributed in $\phi$ direction, and
logarithmically distributed in $r$ direction. The grid numbers ($\nrhd$, $\nphihd$) resolve the gas scale height at the location of the planet with at least 5 grid cells and make the grid cells square shape at the planet location. We also do convergence tests by doubling the resolution, finding that  the gap depth variations are less than $20\%$. So we keep on using the resolution in Table \ref{tab:fg} to minimize simulation time during each $\ite$ step.

Regarding the physical model setup, we only include gas in our simulations without dust and the radial initial gas density profile is $\Sigma_g = \Sigma_{0} (r/r_0)^{-1}$. We assume the whole disk mass is 0.028$\Msun$, which is a normal choice for solar-mass star \citep[e.g., see review in][]{manara_demographics_2022}.
The EoS is assumed locally isothermal and the flared disk is built with aspect ratio $\hpr \propto r^{1 / 4}$. However, since we update $\cs$ for each iteration step as described in section \ref{sec:workflow}, we only use aspect ratios and flaring index as the initial conditions but do not need to use them at any later step of evolution. 
We adopt a \citet{shakura_black_1973} viscosity parameter $\alpha = 10^{-3}$. We use the scale-free parameter setup in $\fargo$ which means $G$, $\Mstar$, $r_0$ = 1. Here we set $r_0 = r_p$ and fix planets at circular orbits. 
The indirect term of potential is included in the simulations.
The planets are introduced into disks from the beginning of the simulations without including any accretion onto the planets. We also examine introducing planets into a disk with a mass-taper function but find no significant difference in the results. 

At the radial boundaries, we adopt power-law extrapolation densities and Keplerian extrapolation azimuthal velocities at both $\rmin$ and $\rmax$.
In terms of radial velocities, we adopt an outflow inner boundary and a symmetric outer boundary. Periodic boundaries are imposed in the azimuthal direction. 

\begin{table}
\caption {$\fargo$ main parameters. Parameters in each column below the second row in this table are corresponding to the cases of planet location $\rp=$ 4, 10, or 30 au, respectively.} \label{tab:fg} 
\centering
\begin{tabular}{llll}
\hline
\hline
parameters      & \multicolumn{3}{c}{values}     \\
\hline
$\Mp$  & \multicolumn{3}{c}{{3$\Mj$, 1$\Mj$, 100$\Me$, 10$\Me$}}       \\
$\alpha$           & \multicolumn{3}{c}{0.001}      \\
\hline
$r_{0}$ = $r_p$ {[}au{]} & 4        & 10       & 30       \\
$r_{min}$ {[}$r_{0}${]}      & 0.25     & 0.1      & 0.033    \\
$r_{max}$ {[}$r_{0}${]}      & 25       & 10       & 3.3      \\
AspectRatio     & 0.04     & 0.05     & 0.066    \\
$\Sigma_{0}[\Mstar/r_{0}^2]$      & $1.8^{-4}$ & $4.5^{-4}$ & $1.34^{-3}$ \\
$N_{r, HD}$               & 580      & 460      & 350      \\
$N_{\phi, HD}$            & 790      & 630      & 480      \\
\hline
\end{tabular}
\end{table}

\subsection{Radiative transfer simulations}
After obtaining the 2D $\sigmag$ and $\cs$ fields from $\fargo$ simulations, we perform 3D Monte Carlo radiative transfer with $\radmc$ \citep{dullemond_radmc-3d_2012}
to obtain the temperature structure. The output gas temperature $\Tgas$ is used to update the corresponding $\cs$ field for $\fargo$ (see Section\,\ref{sec:workflow}). Within $\radmc$ simulations, all parameters are in units of cgs and the main parameters are shown in Table \ref{tab:rad}.

For grid cell setup, $\radmc$ keeps the same global transition disk simulation domain in (r, $\phi$) direction as $\fargo$. The vertical domain is $\theta =  \left[\pi / 2-0.5, \pi / 2\right]$ with mirror symmetry along the midplane. The azimuthal and vertical directions are sampled in linear space, while the radial direction is sampled in logarithmic space. 
We test different combinations of grid resolutions and decide ($\nrrt$, $\nphirt$, $\nthetart$ = (256, 30, 53) in radial, vertical, and azimuthal direction is a proper resolution for using $\nphot = 10^8$ photon packages.
For small $N_{r}, N_{\phi}$, the asymmetry temperature feature in disks due to eccentric gaps is not recovered properly.
For larger $N_{r}, N_{\phi}$, the $\Tmid$ map gets bad photon statistics and it is noisy unless we adopt a larger number of photons $\nphot>10^{9}$, which takes more than 10 hours with paralleling 40 threads for just one iteration step. Also, $\nphot = 10^{8}$ gets similar smooth temperature results as $\nphot>10^{9}$ with more grid cells. Hence, we keep $\nphot = 10^{8}$  for all the simulations presented in this paper.  After $\radmc$, we interpolate the values in $\radmc$ grid cells to match the (r, $\phi$) grid cells in $\fargo$.

For the stellar parameters, we adopt typical values for a T Tauri star, $\Mstar = 1\Msun$, $\Rstar = 1.7\Rsun$, and  $\Tstar = 4730$K.
We only consider stellar radiation as the heating source and ignore viscous heating. 
We assume silicate dust particles with isotropic scattering and the intrinsic density is $3.710\,$g\,cm$^{-3}$.
We also assume dust to gas mass ratio $\epsilon=0.01$ and dust grain size of $0.1\mu m$. As the small dust grains couple well to the gas, we do not assume any dust settling. Also, we do not consider any dust evolution process, such as dynamics, growth, or fragmentation of particles \citep{birnstiel_gas-_2010}. 
The disk density distribution in three dimensions is assumed to be
\begin{equation}
\label{eq:rho}
\rho_{d}(r, z, \phi)=\frac{\Sigma_{d}(r, \phi)}{\sqrt{2 \pi} H(r)} \exp \left(-\frac{z^{2}}{2 H(r)^{2}}\right)
\end{equation}
where $\Sigma_{d}(r, \phi)$ is the dust surface density and $\Sigma_{d}(r, \phi) = \epsilon\Sigma_{g}(r, \phi)$. $H(r)$ is the gas pressure scale height and $
z=r \tan \theta$. 

\begin{table}
\centering
\caption{$\radmc$ parameters.}
\label{tab:rad}
\begin{tabular}{ll}
\hline
\hline
parameters   & values \\
\hline
$\Mstar$ [$\Msun$] & 1      \\
$\Rstar$ [$\Rsun$] & 1.7    \\
$\Tstar$ [K]    & 4730   \\
$N_{photon}$      & $10^8$   \\
$\epsilon$      & 0.01   \\
$N_{r, RT}$            & 256    \\
$N_{\phi, RT}$         & 30     \\
$N_{\theta, RT}$     & 53    \\
\hline
\end{tabular}
\end{table}

%%%%%%%%%%%%%%%%%%%%%%%%%%%
\subsection{Workflow} \label{sec:workflow}

Our iterative approach makes use of $\fargo$ and $\radmc$ codes. 
The workflow of our iteration method is illustrated in Figure\ref{fig:workflow_it}. The green, blue, and yellow boxes represent the steps of $\fargo$, $\radmc$, and post-processing from $\radmc$ to $\fargo$, respectively. Our methodology consists of the following steps:

\textbf{Step 1:} 
We set up our initial physical disk models without planets by assuming azimuthal symmetric 1D gas surface density $\Sigma_{g,0}(r)$ and aspect ratio $\hpr$ of the disks (shown in Box(1)). 
Then, we output $\fargo$ results of 0 orbit to obtain initial 2D $\Sigma_{g,0} (r, \phi$) and sound speed $\cso (r, \phi$) map.
Note that the energy field outputs in $\fargo$ simulations in this paper are actually the isothermal $\cs$. 

\textbf{Step 2:}
2D surface density field from $\fargo$ are read by $\radmc$ and extend to 3D volume density by following Eq.~\ref{eq:rho}, where the scale height $H$ is calculated from $\fargo$ $\cs$ field.
Then the dust radiative transfer simulations are conducted (Box (2)), and the output of the dust temperature $\Tdust(r, \theta, \phi)$ is obtained. As $\radmc$ does not include any  photochemistry simulations, we assume  $\Tgas(r, \theta, \phi)) = \Tdust(r, \theta, \phi)$. From $\Tdust(r, \theta, \phi)$, the midplane temperature $\Tmid$(r, $\phi$) can be obtained (Box (3)). 
Using this $\radmc$ temperature as the non-planet disk temperature can help us to get rid of the initial temperature profile assumption in $\fargo$. In fact, such a step is also done in Fig. 6(a) in \citet{bae_ideal_2019} to get the first Monte Carlo radiative transfer (MCRT) temperature, which aims to get rid of the assumed stellar irradiation-dominated
temperature T$_{\rm{irr}}$. We also test the iteration process (assuming vertical hydrostatic equilibrium ) described in Appendix A in \citet{bae_ideal_2019} to get the multiple iteration MCRT temperature but the differences between the first MCRT temperature and multi-time MCRT temperature in our disk model are negligible. This MCRT iteration process makes no difference in our case but at least doubles our MCRT workload and costs much more computation time. So we directly use our $\radmc$ temperature for later steps. 

\textbf{Step 3:}
By using the $\Tmid$(r, $\phi$) from the last step, we could infer a new $\cs$ field by assuming a vertical isothermal approximation. 
Even though we still use the isothermal assumption here, because of the non-smooth $\Tmid$(r, $\phi$) reflecting the gap-opening process, such new $\cs$ does not equal the initial isothermal $\cso$ anymore. The new $\cs$ is treated as the new gasenergy.dat file for the next $\fargo$ run (Box (4)). This is the important step that moves beyond the isothermal assumption in the conventional non-iteration method and shows the feedback effect of the gap-opening process. 
In Appendix \ref{app:Tmid_vs_Tavgz}, we have a test to compare a vertical density weighted temperature with $\Tmid$. We find that they are similar, especially in gap regions. For simplicity, we use $\Tmid$ in this paper.
Next, we restart the $\fargo$ simulation and evolve it over 100 orbital times (we assume the iteration step is 100 orbit here) (Box(5)) and as a result, we get the output as Box(6). Again, during the $\fargo$ step, the EoS is assumed isothermal. \citet{malygin_efficiency_2017, pfeil_mapping_2019} demonstrate that the thermal relaxation time varies across the disk, and in some regions, there is large cooling time ($>$100 local orbits), where our iteration time is a good approximation. However, in some outer disk regions, like a few tens of au, (the specific regions depend on the model conditions) have short cooling time, where our choice of 100 orbits can be too long. As a test, we performed simulations with iteration steps of 50 orbits in Appendix \ref{app:iteration steps}, and found no difference with the 100 orbits case. We also test $\ite$ step of 100 orbits against 500 orbits in Appendix \ref{app:iteration steps}, which do not converge very well in gap regions. It means the iteration step of 500 orbits could not replace 100 orbits. For these reasons, we keep 100 orbits for all the main simulations of this work.

\textbf{Step 4:}
The result of 100 orbit $\fargo$ is used as the input for $\radmc$ (Box(7)). During the $\radmc$ setup, the gas pressure scale height $H$ is given by the $\fargo$ $\cs$, $H = \cs/\Omega$. Therefore, the extension of 2D $\sigmag$ to 3D volume density $\rho$ (shown as Equation \ref{eq:rho}) can be also modified by the gap-opening feedback.

\textbf{Step 5:}
Repeat Step 2 to Step 4 and iterate until reaching a quasi-steady state, which also means the iteration process is from Box(2) to Box(7). We iterate all the simulations over 2000 planet orbital time which corresponds to $1.6\times10^{4}$ yrs for $\rp$ = 4 au, $6.4\times10^{4}$ yrs for $\rp$ = 10 au, $3.3\times10^{5}$ yrs for $\rp$ = 30 au, respectively.

In summary, there is density and velocity evolution but no energy/$\cs$ evolution over time in HD simulations, while the energy/$\cs$ field is evolved by executing RT simulations. Meanwhile, the evolving $\cs$ field contains the information from the gap-opening feedback. As a comparison, in this paper, the conventional $\noit$ method is running $\fargo$ then $\radmc$ simulation once. To be more specific, $\noit$ uses $\fargo$ with the physical assumptions (initial isothermal $\cs$) in Box(1) to obtain $\sigmag$. Then input this $\sigmag$ into $\radmc$ to get the temperature $\Tmid$. The whole process is finished after doing this once. 

%%%%%%%%%%%%%%%%%%%%%%%%%%%%%%%

\section{Results}
\label{sec:result}

\begin{figure*}
\includegraphics[width=\linewidth]{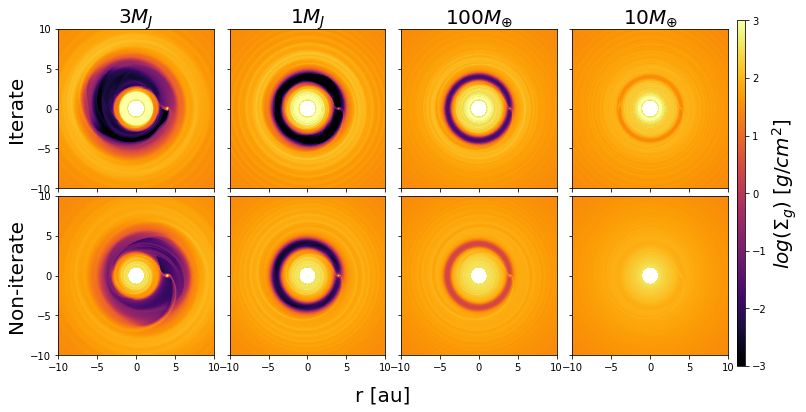} 
\centering
\caption{2D gas density map of planets at fixed radius of 4 au over 2000 orbits of iteration method (upper panels) and non-iteration method (lower panels). From left to right columns, gaps are opened by planets of $3\Mj$, $1\Mj$, $100\Me$, and $10\Me$, respectively.
}
\label{fig:dens4au} 
\end{figure*}

In this section, we describe the results of our simulations, and  compare the results between the $\ite$ and $\noit$ methods.

%%%%%%%%%%
\subsection{Gas surface density}

Based on Step 3 in the iteration workflow described in section \ref{sec:workflow}, we can obtain the surface density in disks.
From left to right columns, Figure \ref{fig:dens4au} shows the 2D gas surface density maps of gaps opened by planets at 2000 orbits in masses of 3$\Mj$, 1$\Mj$, 100$\Me$, and 10$\Me$ at orbital radii of 4 au. The $\ite$ and $\noit$ results are presented in upper and lower panels, respectively.
The gaps from the $\ite$ method are generally deeper and wider than their counterparts simulated by the $\noit$ method. 

As gap structures are shown in most simulations, we quantify the gap width and depth from the data of surface density and compare $\ite$ with the $\noit$ models. In this work, we define the gap width $\gw$ with the method in \citet{kanagawa_mass_2016} which is the radial region where $\gd \leq 0.5$. Meanwhile, we define the gap depth $\gd$ as that in \citet{fung_how_2014} which is the radial averaging value within $2 \times max(\rhill, H)$ of the planet, where $\rhill$ and $H$ are hill radius and scale height at $\rp$. Both the gap width and gap depth are obtained by azimuthal averaging and the last 500-orbit averaging.

\begin{figure}
\includegraphics[width=\columnwidth]{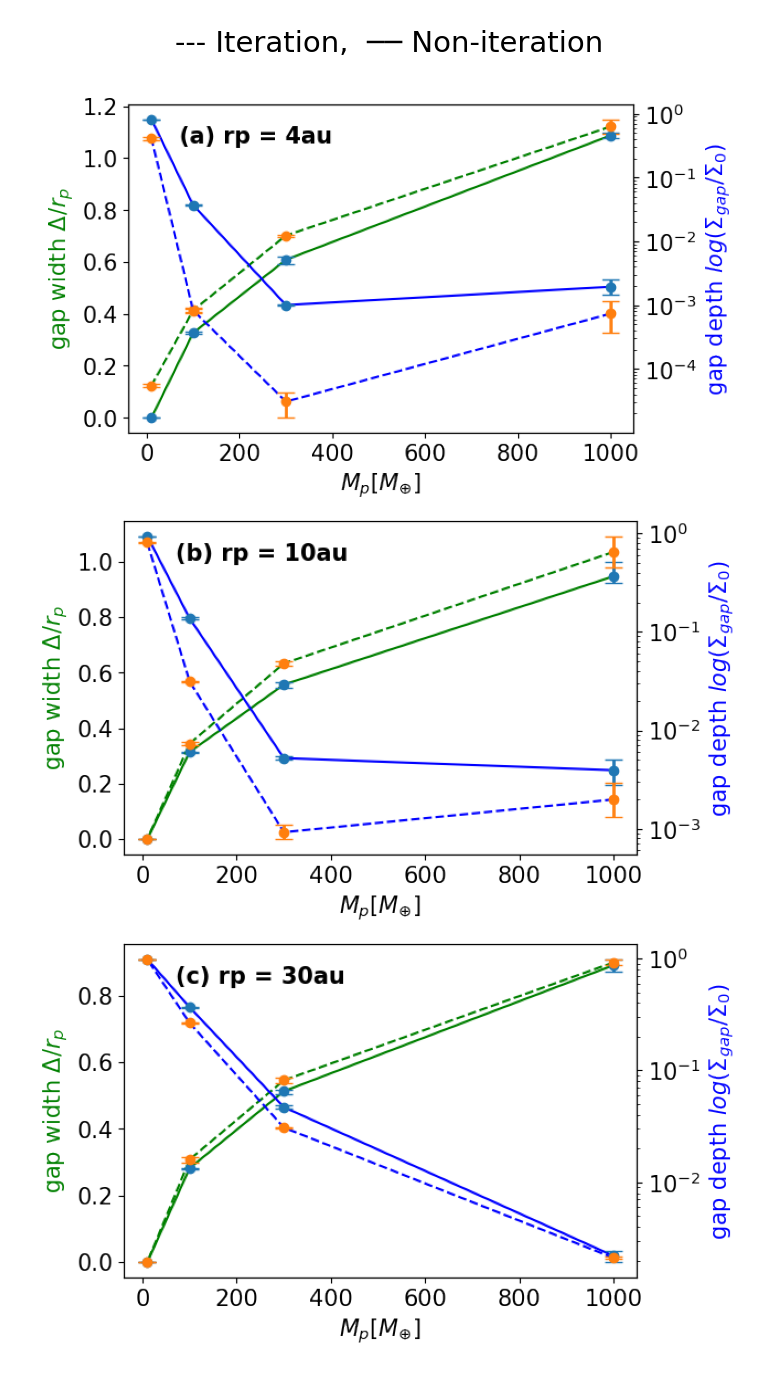} 
\caption{Normalized gap widths $\gwn$ (green) and depths $\gd$ (blue) and their uncertainties (error bars) as a function of $\Mp$ at $\rp$ = 4 (panel (a)), 10 (panel (b)), and 30 au (panel (c)), respectively. Dashed and solid lines represent the $\ite$ and $\noit$ results. 
}
\label{fig:gap} 
\end{figure}

Figure \ref{fig:gap} displays the comparison of the normalized gap width $\gwn$ (green) and gap depth $\gd$ (blue) as a function of $\Mp$ of $\ite$ (dashed lines) and $\noit$ (solid lines) methods. From top to bottom panels, $\rp$ = 4, 10, and 30 au, respectively. Uncertainty of gap depths and widths are also shown, which come from time averaging of the last 500 orbits. The choice of final 500 orbits is because our simulations appear to reach quasi-steady states at around 1500 orbits though gap depth and gap eccentricity are still slightly fluctuating.
Under the definition of gap width and gap depth here, $\Mp=10\Me$ cannot open gaps except for $\rp$ = 4 au of $\ite$. With regard to the $\gd$ of $\noit$ method, we find them consistent with the empirical formulas in \citet{fung_how_2014} except the eccentric case caused by $\Mp=3\Mj$. 
In general, the iteration method infers a slightly wider gap width $\gw$ than the $\noit$ method, whereas $\ite$ predicts an order of magnitude deeper gap depth $\gd$ than $\noit$. 
The reason for the deeper gap in $\ite$ is the aspect ratio $\hpr$ at the gap region is smaller than the $\noit$. Based on equation 3 in \citet{fung_how_2014}, a lower $\hpr$ lead to a smaller $\sigmagap$. As the $\ite$ method predicts a deeper gap than the $\noit$ method, which means a lower mass planet can possibly open a deep gap. 
For instance, in Figure \ref{fig:gap}(a), when $\rp$ = 4au, the $\ite$ predicts that a Saturn mass planet can open a gap as deep as a Jupiter mass planet in the $\noit$ method. 
This can help to explain why massive planets predicted by usual $\noit$ simulations are supposed to be observable but have not actually been widely detected in real observations.

Among the gap depth of the $\ite$ method, as $\Mp$ increases, $\gd$ decreases, though this trend is invalidated to $\Mp = 3 \Mj$ at $\rp = 4$ or 10 au. In these two cases, planets open appreciable eccentric gaps and streamer structures appear, which increases gas density in gaps. Similar situations also happen in the $\noit$ cases, though the streamers are less strong and the measured $\gd$ are close for $\Mp = 1\Mj$ and $3\Mj$. 
In terms of gap width $\gw$, as $\Mp$ increases, $\gw$ increases. For $\Mp = 3\Mj$, it can open a gap roughly as wide as the planet orbit $\rp$ in our disk models. 
For a fixed $\Mp$, if $\rp$ increases, the normalized gap width $\gwn$ is smaller and the gap depth $\gd$ is shallower. 
This is because the higher disk scale height $\hpr$ in the outer disk makes pressure torque stronger to prevent the gap opening process. 
Besides the disk density profiles of planets at 4 au, Fig.~\ref{fig:dens10au} and \ref{fig:dens30au} in Appendix 
show the 2D gas surface density map of planets at 10 au and 30 au. As planets move further away from the central stars, they open shallower gaps than their counterparts at 4 au.

%%%%%%%%%%
\subsection{Midplane temperature} \label{sec: Tmid}

After implementing dust $\rt$ (described in Step 2 in section \ref{sec:workflow}) and assuming $\Tgas = \Tdust$, we get the 3D $\Tgas$ structure of disks. As we are concerned about 
icy-pebbles or planetesimals which mainly concentrate at the disk midplane, we focus on the midplane temperature $\Tmid$ derived from both $\ite$ and $\noit$ methods.

\begin{figure*}
\includegraphics[width=\linewidth]{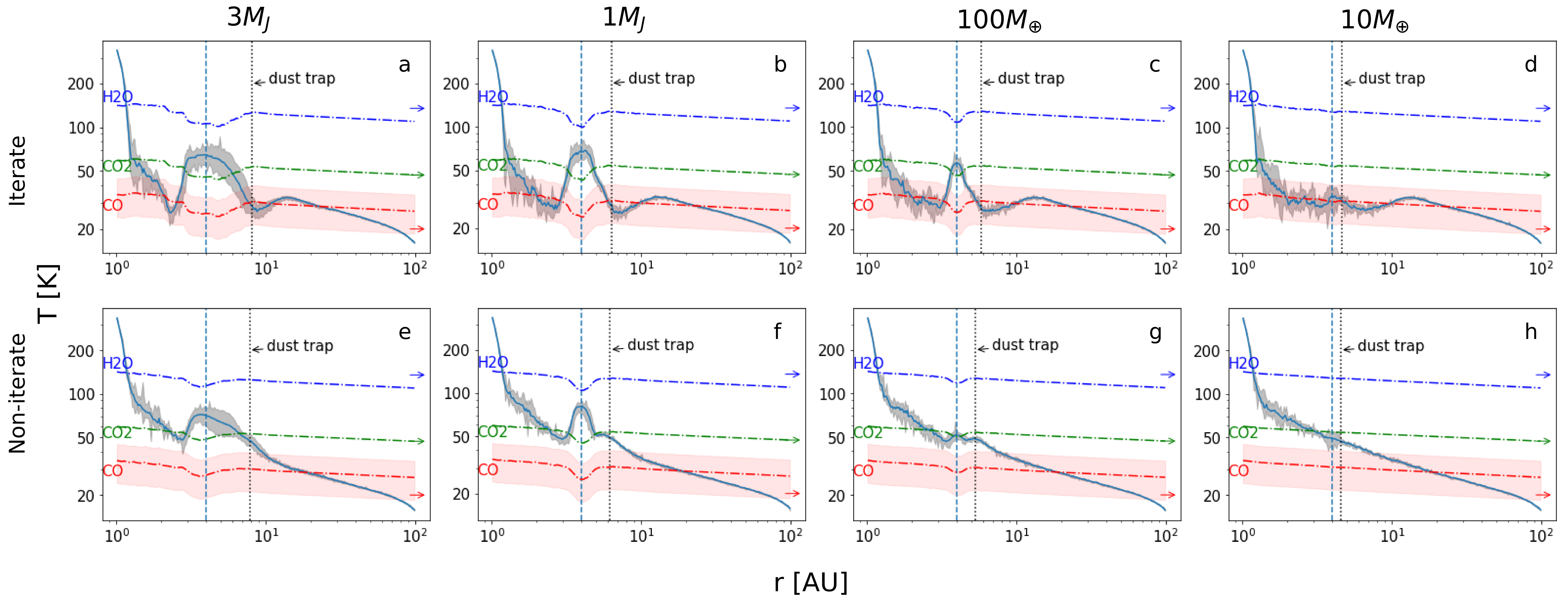} 
\caption{Midplane temperature $\Tmid$ as a function of disk radius from $\radmc$ simulations of planets at 4 au over 2000 orbits of $\ite$ method (upper panels) and $\noit$ method (lower panels). The cyan solid lines represent the azimuthal averaged $\Tmid$, while the shading areas represent the deviation of the profiles along different azimuthal angles. From left to right, there are results of $3\Mj$, $1\Mj$, $100\Me$, and $10\Me$, respectively. The blue, green, and red dash-dotted lines near the horizontal direction represent the pressure-dependent sublimation temperature $\Twater$ $\Tcd$, and $\Tcm$, respectively. $\Tcm$ is shading with a light red region to highlight the wide range of possible values calculated from different binding energies given by KIDA. As a comparison, constant $\Tsub$ in \citet{oberg_effects_2011} are marked with short arrows in these three colors on the right edge of each panel.
The vertical cyan dashed lines and grey dotted lines mark the location of the planets and pressure maximum/dust trapping. 
}
\label{fig:T4au} 
\end{figure*}

Figure \ref{fig:T4au} shows the comparisons between azimuthal averaged $\Tmid$ (cyan lines) of planets at 4 au over 2000 orbits calculated by $\ite$ method (upper panels) and $\noit$ method (lower panels). The gray shading regions represent the $\Tmid$ in different azimuthal angles.
Three molecules and their pressure-dependent sublimation temperatures $\Twater$, $\Tcd$, and $\Tcm $ (calculations follow the recipe in \citet{hollenbach_water_2009}) are marked as blue, green, and red dash-dotted lines, respectively. We use binding energy provided on KIDA\footnote{https://kida.astrochem-tools.org} database. The uncertainty of $\Tcm$ due to different binding energy choices is shown as a light red shading area.
As for comparison, the constant sublimation temperatures $\Twater$ = 125K, $\Tcd$ = 47K, $\Tcm $ = 25K in \citet{oberg_effects_2011} are marked as short horizontal arrows in corresponding colors. 
Overall, the $\ite$ method predicts distinct $\Tmid$ when compared with the $\noit$ method at two regions. In particular, in the inner disk regions(r < 10au) and the gap regions. 

At the inner disk, iterated $\Tmid$ drops more rapidly than non-iterated $\Tmid$. For example, we can clearly see the differences between Panel d and h in these three figures. As $\Mp = 10\Me$, such low planet has negligible effects on disk temperature as they are difficult to open gaps to influence $\Tmid$. 
Therefore, the difference between Panel d and h does not come from the planet opening gaps, instead, the difference comes from the methods we adopt, $\ite$ or $\noit$. 
The underlying physical explanation will be discussed in more detail in section \ref{sec: T drop}. In short, the puff-up of the scale height at the inner dust rim cause a strong shadowing effect and lower the temperature in these regions.

At the gap regions, $\ite$ predicts more significant $\Tmid$ contrasts between inside gap regions and outside gap edges than the \noit. The highest contrast of $\Tmid$ can be up to 
40K (increase from 30K to 70K) when $3\Mj$ or $1\Mj$ at 4au of $\ite$ method (see Panel a and b). 
The underlying explanation is the $\ite$ tends to open deeper gaps than the $\noit$ and allows more stellar photons to penetrate into the midplane and increase $\Tmid$. 
However, the peak values of $\Tmid$ from both methods are similar in the same $\Mp$ and $\rp$ conditions. 

Regarding the $\ite$ results, as $\Mp$ increases, the $\Tmid$ at gaps increases more significantly. It is because more massive planets are able to open deeper and wider gaps and more stellar photons can penetrate deeper at the gap region and heat up midplane dust and gas. Such a trend is also seen in the $\noit$ method. Furthermore, the midplane temperature of $\rp = 10, 30$ au are shown in Figure \ref{fig:T10au} and \ref{fig:T30au}, respectively.

\begin{figure}
\includegraphics[width=0.8\columnwidth]{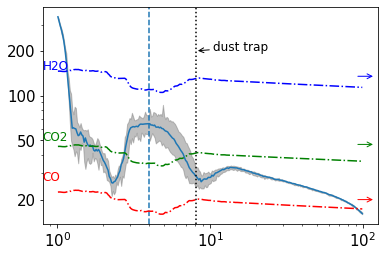} 
\centering
\caption{Similar to Figure \ref{fig:T4au} (a) but $\Tsub$ is calculated by using molecule binding energy in \citet{oberg_effects_2011}. 
}
\label{fig:Tsub_be} 
\end{figure}

By combing the sublimation temperature and the disk midplane temperature, we can measure where the midplane icelines of different molecules are in section \ref{sec: icelines}. 
The numbers of icelines primarily rely on the values of $\Tsub$ and the disk $\Tmid$. 
If we use the values of binding energy suggested in \citet{oberg_effects_2011}, the overall profiles of sublimation temperature of all these three volatile will shift up or down. Figure \ref{fig:Tsub_be} is a plot of $\Tmid$ but with $\Tsub$ calculated from binding energy adopted by \citet{oberg_effects_2011}. Compared with \ref{fig:T4au}(a), now the whole $\Tcm$ shifts lower significantly and the CO iceline moves outward dramatically to around 90au, and only one CO iceline exists. 
Therefore, in this case, particles or pebbles with \ce{CO} ices only exist in the very outer disk.

%%%%%%%%%%%%%%%%%%%%%%%%%%%%%%%%%%%

\subsection{Eccentricity} 

When comparing the density results within either $\ite$ or $\noit$ method, if the planet masses $\Mp < 1\Mj$, they open quite circular gaps and more massive planets open deeper and wider gaps. 
In terms of the most massive cases of $3\Mj$ in our modeling, the planets open eccentric gaps. Our results agree with the results from \citet{kley_disk_2006} who found that planets with mass $\Mp > 3\Mj$ open eccentric gaps in disks with a viscosity of $\nu = 10^{-5}$ or $\alpha \approx 0.004$. In this section,  we quantify the eccentricity $e$ of the gaps opened by $\Mp = 3\Mj$ or $1\Mj$ with two kinds of methods. Because the inner and outer edge of a gap has different eccentricities, we measure them separately. 

\begin{figure}
\includegraphics[width=\columnwidth]{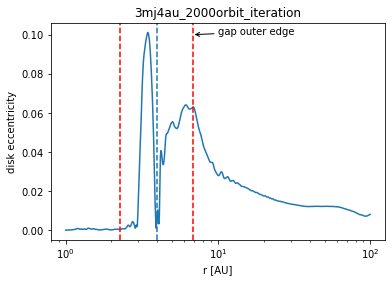} 
\caption{Disk eccentricity calculated from Equation \ref{eq:ecc} for an example of $3\Mj$ at 4au of $\ite$ method. Planet location is marked with a vertical blue dashed line. The gap inner and outer edges are marked by red dashed lines.}
\label{fig:ecc_formula} 
\end{figure}

The first method is obtaining  $e$ by fitting ellipses to the shape of the inner/outer edges of gaps. The second method is using equation 28 in \citet{ju_global_2016}
\begin{equation}
e(r)=\frac{\left|\int d \phi \Sigma(r, \phi) v_{r} \exp (i \phi)\right|}{\int d \phi \Sigma(r, \phi) v_{\phi}}
\label{eq:ecc}
\end{equation}
to calculate $e$ at the location of the inner/outer edges of gaps. Figure \ref{fig:ecc_formula} displays an example of using the equation \ref{eq:ecc} to calculate the $e$ as a function of radius of the simulation of $3\Mj$ at 4 au over 2000 orbit by using $\ite$ method. At this case, $e \sim 0.06$ at the gap outer edge, which is not very different from the value in \citet{kley_disk_2006} though the disk parameters (e.g. viscosity $\alpha$, aspect ratio $H/r$) are not exactly the same. 

Table \ref{tab:ecc} summarizes gap eccentricities $e$ from different simulations measured by different methods, fitting ellipse in "graph" or calculating with "formula". Names of different cases are written in abbreviations. For example, 3mj4au\_it\_graph means the case of 3$\Mj$ at 4 au of $\ite$ measured by graph method, and so on. 
In both methods, we average the values of the last 500 orbits (shown as "mean" in Table \ref{tab:ecc}) and calculate their standard deviations (shown as "std" in Table \ref{tab:ecc}). In general, both $\ite$ and $\noit$ methods get similar $e$. Also, the values of $e$ are closed from the graph and formula measuring method. $\Mp = 3\Mj$ induces relatively high $e \sim 0.07$ when $\rp = $ 4 or 10 au, which could also be seen from the eccentric gaps in Figure \ref{fig:dens4au} and Figure \ref{fig:dens10au}. $\Mp = 1\Mj$ only open gaps in almost circular shapes. 

The high eccentricity of a gap can have a non-neglectable effect on the temperature. For the most eccentric case, for example, $e=0.08$ measured by graph fitting of 3mj4au of the $\ite$ method, we can obtain $r_{min}$= 6 au and $r_{max}$ = 7.6 au from fitting the ellipse of the outer gap edge. By plugging them into the corresponding temperature profile, we can find $\Tmid$ varies from about 40 to 27K. In other words, CO ice might exist at the semi-major axis side but sublimate at the semi-minor axis side of the outer edge of gaps.

\begin{table}
	\centering
	\caption{Comparison of inner and outer gap edge eccentricity $e$ of massive planets at different $\rp$ from $\ite$ or $\noit$ method. $e$ is measured by either "graph" or "formula" method and the averaged values of the last 500 orbits are marked as "mean" and the uncertainty are marked as "std". Values of $e>0.05$ are highlighted with purple background.}
	\label{tab:ecc}
 \resizebox{\columnwidth}{!}{%
\begin{tabular}{lllll}
\hline
\hline
                       & e\_in\_mean & e\_in\_std & e\_out\_mean & e\_out\_std \\ \hline
3mj4au\_it\_graph      & 0.01        & 0          & \cellcolor{blue!25}0.08         & 0.03        \\
3mj4au\_it\_formula    & 0           & 0          & \cellcolor{blue!25}0.06         & 0           \\
1mj4au\_it\_graph      & 0.01        & 0          & 0.01         & 0.01        \\
1mj4au\_it\_formula    & 0           & 0          & 0.01         & 0           \\
3mj4au\_nonit\_graph    & 0           & 0          & \cellcolor{blue!25}0.08         & 0.04        \\
3mj4au\_nonit\_formula  & 0           & 0          & \cellcolor{blue!25}0.07         & 0           \\
1mj4au\_nonit\_graph    & 0.02        & 0          & 0.01         & 0           \\
1mj4au\_nonit\_formula  & 0           & 0          & 0.01         & 0           \\ \hline
3mj10au\_it\_graph     & 0.01        & 0          & \cellcolor{blue!25}0.06         & 0.02        \\
3mj10au\_it\_formula   & 0           & 0          & \cellcolor{blue!25}0.07         & 0           \\
1mj10au\_it\_graph     & 0.01        & 0          & 0.01         & 0           \\
1mj10au\_it\_formula   & 0           & 0          & 0.01         & 0           \\
3mj10au\_nonit\_graph   & 0.03        & 0.02       & \cellcolor{blue!25}0.06         & 0.03        \\
3mj10au\_nonit\_formula & 0.02        & 0          & \cellcolor{blue!25}0.07         & 0.01        \\
1mj10au\_nonit\_graph   & 0.04        & 0.01       & 0            & 0           \\
1mj10au\_nonit\_formula & 0.01        & 0          & 0            & 0           \\ \hline
3mj30au\_it\_graph     & 0.01        & 0          & 0.01         & 0           \\
3mj30au\_it\_formula   & 0           & 0          & 0            & 0           \\
1mj30au\_it\_graph     & 0           & 0          & 0            & 0           \\
1mj30au\_it\_formula   & 0           & 0          & 0            & 0           \\
3mj30au\_nonit\_graph   & \cellcolor{blue!25}0.05        & 0.03       & 0            & 0           \\
3mj30au\_nonit\_formula & 0.04        & 0          & 0            & 0           \\
1mj30au\_nonit\_graph   & 0           & 0          & 0.01         & 0           \\
1mj30au\_nonit\_formula & 0           & 0          & 0            & 0           \\ \hline
\end{tabular}%
}
\end{table}

\subsection{Dust trap} 

Dust grains in a disk experience radial drift unless they get trapped in gas pressure bumps (e.g. \citet{pinilla_ring_2012, pinilla_trapping_2012, dullemond_disk_2018}). We can investigate pressure gradient profiles around the gaps to see how our model affects dust-trapping conditions. Pressure is given by $
P(r)=\Sigma(r) c_{\mathrm{s}}^{2}(r)
$. Here both the gas surface density and sound speed $\cs$ are azimuthally averaged after 2000 orbits. 
Different pressure gradients from $\ite$ and $\noit$ can lead to different efficiency of dust trapping. Thus, dust of different sizes could be distributed differently. 

\begin{figure}
\includegraphics[width=\columnwidth]{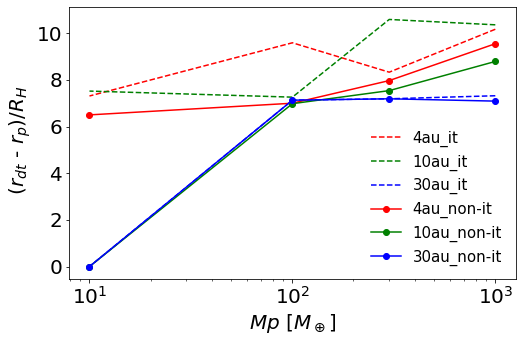} 
\centering
\caption{Normalized dust trapping location as a function of planet mass $\Mp$. The $\ite$ and $\noit$ are marked with dashed and solid lines and different $\rp$ = 4, 10, and 30 au are in red, green, and blue, respectively.
}
\label{fig:rdt} 
\end{figure}

The location of the pressure maximum, also named dust-trapping location $\rdt$ here, is when the pressure gradient is zero ($\dlogpr = 0$). We obtain the $\rdt$ of different cases of $\Mp$ and $\rp$ of the $\ite$ and $\noit$ methods. We find that there are no dust traps in the cases of $\Mp$ = 10 $\Me$ when planets are at $\rp$ =10 au of the $\noit$ method, and $\rp$ = 30 au of both methods.
Figure \ref{fig:rdt} shows the normalized dust trapping location $(\rdt - \rp)/ \rhill $ as a function of planet mass $\Mp$. If for a specific case, there is no $\dlogpr=0$, we put $(\rdt - \rp)/ \rhill = 0$.
In general, if a planet can form a pressure maximum to trap dust around the outer gap edge, $\rdt - \rp = 7 \sim 10 \rhill$ regardless of different $\Mp$ and $\rp$. The dust-trapping locations $\rdt$ from both methods do not show a big difference, especially for $\rp$ = 30 au cases. In smaller $\rp$ cases, $\ite$ tends to trap grains in slightly outer locations than $\noit$ for a given $\Mp$ and $\rp$. 
Furthermore, $\rdt-\rp$ is roughly equal to the gap widths in each case. In other words, the outer gap edges are approximately the middle points between planets and dust-trapping locations. 

As small dust particles can couple well with gas and may flow through the dust trap, there should be a minimum grain size so that grains larger than this threshold can be trapped by the pressure bump. Thus, we can further infer what ratio of dust is trapped by a pressure bump. The minimum particle size that can be trapped is described by: \citet{pinilla_ring_2012} 

\begin{equation}
a_{\text {critical }}=\frac{6 \alpha \Sigma_{\mathrm{g}}}{\rho_{\mathrm{s}} \pi|(\mathrm{d} \log P / \mathrm{d} \log r)|}\left|\left(\frac{3}{2}+\frac{\mathrm{d} \log \Sigma_{\mathrm{g}}}{\mathrm{d} \log r}\right)\right|
\end{equation}

In our modeling, $\alpha=10^{-3}$, $\rho_s = 3.710 g/cm^3$. We find that the $\ite$ has similar $\acri$ of about 0.3cm as the $\noit$ results at the location of pressure maximum $\rdt$ . 
Furthermore, we assume grain size distribution follows \citet{mathis_size_1977}, $n(a) \propto a^{-p}$, where $p = 3.5$. 
The range of dust sizes is from $0.1 \mu m$ to $\afrag$, where $\afrag$ is the maximum particle size before they fragment due to turbulent relative velocities $\vf$ \citep{birnstiel_simple_2012}:
\begin{equation}
a_{\mathrm{frag}}=\frac{2}{3 \pi} \frac{\Sigma_{\mathrm{g}}}{\rho_{\mathrm{s}} \alpha} \frac{v_{f}^{2}}{c_{\mathrm{s}}^{2}}
\end{equation}

Follow \citet{pinilla_ring_2012}, we set $\vf$ = 10\,m\,s$^{-1}$. We find that the $\ite$, $\afrag$ could be a few times higher than $\noit$. By using the number of particles at a size beam $nda$ times the mass of a particle(assume spherical particle) and integrating from $\acri$ to $\afrag$, we can obtain the fraction of dust mass that gets trapped in the dust trapping regions discussed. The dust-trap fraction is:
$f_{trap} = (a_{frag}^{0.5}-a_{cri}^{0.5})/({a_{frag}^{0.5}-a_{\min }^{0.5}})$. $\ftrap$ represents the fraction of the dust mass that could be trapped in the pressure bump region compared to the total dust mass of the pressure bump region. 

If we take the $3\Mj$ at 4$\rp$ as an example, at the region of the pressure bump (assume from outer gap edge $\rgapo$ to $2\rdt - \rgapo$), both $\ite$ and $\noit$ have similar averaged $\acri = 2\times 10^{-2}$ cm. If we assume $a_{min} = 10^{-5}$ cm, and plug in averaged $\afrag = 10$ cm for $\ite$ or averaged $\afrag=3$ cm for $\noit$. Therefore, we have $\ftrap$ = 0.96 for $\ite$ and $\ftrap$ = 0.92 $\noit$, which means a slightly higher fraction of dust mass could be trapped in the pressure bump predicted by the $\ite$. Proper inclusion of dust evolution is needed to test this hypothesis in the future.

\subsection{Distribution of ice species} 
\label{sec: icelines}

Figure \ref{fig:T4au}, \ref{fig:T10au}, and \ref{fig:T30au} show $\Tgas$ from the $\ite$ method and how it deviates from the $\noit$ approach. For this reason, the ices of \ce{H2O}, \ce{CO2}, and \ce{CO} are distributed in different locations when comparing the two methods. More ice means more solid masses could contribute to the pebble or planetesimal formation, while the available species of ice can affect the final planetesimal composition. Therefore, in this section, we use temperature and pressure information to show where different icelines are, where ice distributes throughout the whole disk, and what kinds of ice species form at dust trapping locations. 

For each specific molecular iceline, the number of radial icelines depends on the number of intersections between the disk midplane temperature $\Tmid$ and the radial pressure-dependent sublimation temperature $\Tsub$. 
We notice the bonding energy of \ce{CO} has a wide range of values in KIDA and we show the uncertainty of the $\Tcm$ with light red shading regions in Figure \ref{fig:T4au}, \ref{fig:T10au}, and \ref{fig:T30au}. Therefore, we need to keep in mind that the numbers and locations of \ce{CO} iceline could vary due to adopting different bonding energy of \ce{CO}.

\begin{figure*}
\includegraphics[width=\linewidth]{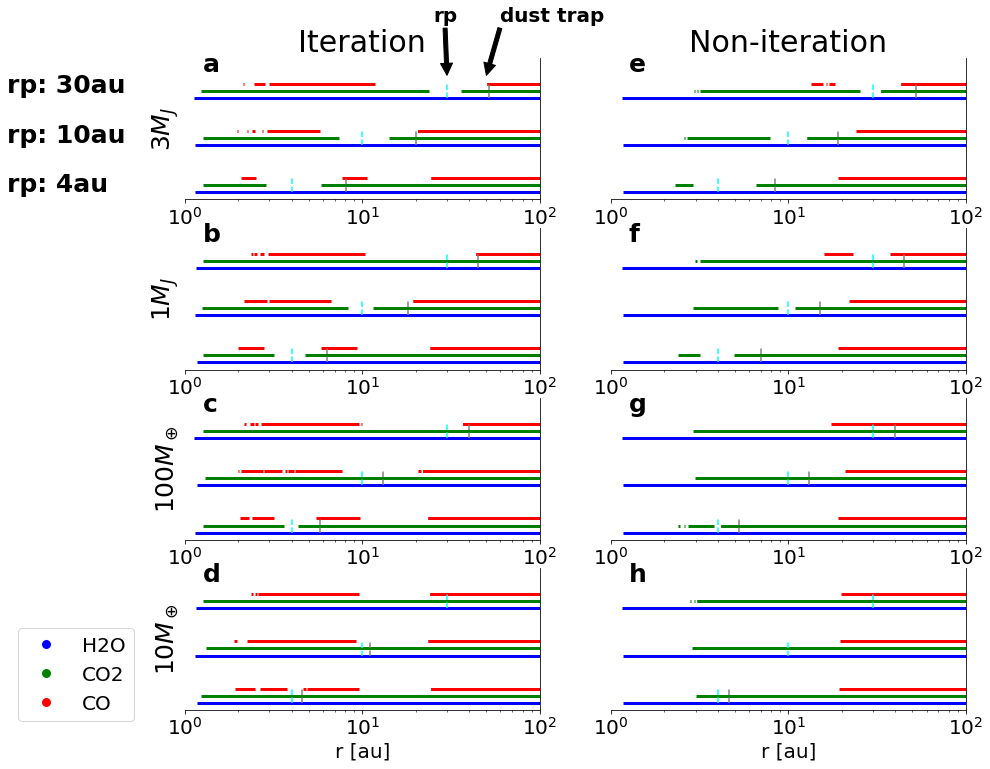} 
\centering
\caption{Ice distribution in disks with different $\Mp$ and $\rp$. The $\ite$ and $\noit$ results are shown in the left and right panels, respectively. High to low $\Mp$ are listed from top to bottom panels. In each panel, for instance, $3\Mj$ of $\ite$ method, there are three groups of data representing the cases of planets at 4 (bottom), 10 (middle), and 30 au (top), respectively. The location of \ce{H2O}, \ce{CO2}, and \ce{CO} ice is in blue, green, and red bars. Vertical cyan dashed lines mark $\rp$ and grey dashed lines display the corresponding dust-trap location.  Note that in some cases there is no grey-dashed line because there is not dust trapping.
}
\label{fig:ice} 
\end{figure*}

Figure \ref{fig:ice}  summarizes the ice distribution of \ce{H2O}, \ce{CO2}, and \ce{CO} throughout the whole disk. The left and right columns show $\ite$ and $\noit$ results. In each column, from top to bottom panels, $\Mp = 3\Mj, 1\Mj, 100\Me$, and $10\Me$ are shown. In each panel, from bottom to top, cases of $\rp$ = 4, 10, and 30 au are shown. The ices of \ce{H2O}, \ce{CO2}, and \ce{CO} are displayed as horizontal blue, green, and red bars, respectively. Planet and dust trap locations are marked as cyan and grey dashed lines. To simplify, here we only consider the azimuthal averaged $\Tmid$ to obtain the radial midplane iceline locations for different volatiles. However, for some high eccentric cases caused by massive planets, the iceline locations can vary at different azimuthal angles. In addition, we define an iceline as the boundary where volatile freeze out and condense into solid, but do not count it when volatile sublimates.

Without considering the time evolution and dust drift, our static ice distribution model gives the following results for the main species:

\ce{H2O} ice: 
all modeling results from the $\ite$ or $\noit$ methods for different $\Mp$ and $\rp$ yield only one water iceline in the disk, and similar iceline locations at around 1.2 au in our disk model. Therefore, planetesimals formed outside 1.2 au can have \ce{H2O} ice.

\ce{CO2} ice: 
Both $\ite$ and $\noit$ methods in the massive planet cases can have two obvious \ce{CO2} icelines shown in Panels a, b, e, and f (except $\rp = $ 30 au in Panels b and f) in Figure \ref{fig:ice}. Because the presence of a massive planet ($\Mp \geq 1\Mj$) opens deep enough gaps that increase the $\Tmid$, which causes \ce{CO2} ice to sublimate at the gap regions and freeze out again in the outer disks. Locations of outer \ce{CO2} icelines are close to gap outer edges. 
In the inner disk, locations of \ce{CO2} icelines are predicted to be around 1.5 au in $\ite$ whereas about 3 au in $\noit$. 

\ce{CO} ice:
One of the most distinct features between the two kinds of models is that the $\ite$ predicts more complicated \ce{CO} icelines features than $\noit$. 
Because of the $\Tmid$ increase at the gap regions, all models of $\ite$ suggest the \ce{CO} ice would sublimate except $10\Mp$ cases. 
For $\rp=4$ au, despite neglecting the short discontinuations in red bars (due to noise in $\rt$ temperature) in those $\ite$ panels, we find three icelines of \ce{CO} in $\ite$ while only one iceline in $\noit$. 
In this case, the \ce{CO} ice can exist in three discrete radial regions in $\ite$ results. The first region, from 1 au to somewhere close to the inner edges of gaps, where the inner disk $\Tmid$ drops. The second region,  from somewhere near the outer edges of the gap to around 10 au, is ascribed to gap heating with the shadowing effect in the inner disk causing $\Tmid$ to drop. The third region starting from about 25 au is due to the outer disk temperature decrease, which is broadly similar to the $\noit$ \ce{CO} ice distribution region outside 20 au.

\section{Discussion}\label{sec:discussion}

Our coupled treatment of hydrodynamics and radiative transfer allows us to shed new light on the feedback of gap-opening planets on the temperature and pressure structure of the protoplanetary disk, which in turn may influence the composition of planetesimals and planets. We discuss this below, followed by a discussion of iceline and disk substructure, as well as caveats of this work and potential improvements for future models.

\subsection{The C/O ratio as a planet formation tracer} 

The carbon-to-oxygen ratio is a potential signature of the history of planet formation \citep{oberg_effects_2011}. Physical and chemical models of protoplanetary disks, with varying degrees of complexity, have been developed to understand the radial behaviour of the C/O ratio \citep[e.g.,][]{cleeves_constraining_2018, zhang_systematic_2019, miotello_bright_2019, bosman_molecules_2021-1}. Recently, an azimuthal C/O ratio variation in $\ppd$ has also been reported and modelled \citep{keyte_azimuthal_2023}. As we have shown, planet-induced gaps introduce significant new complexity to this picture by creating feedback and altering the thermal structure. This in turn modifies various ice lines and thus the C/O-ratio imprinted on subsequently forming planetesimals and planets.

\begin{figure*}
\includegraphics[width=\linewidth]{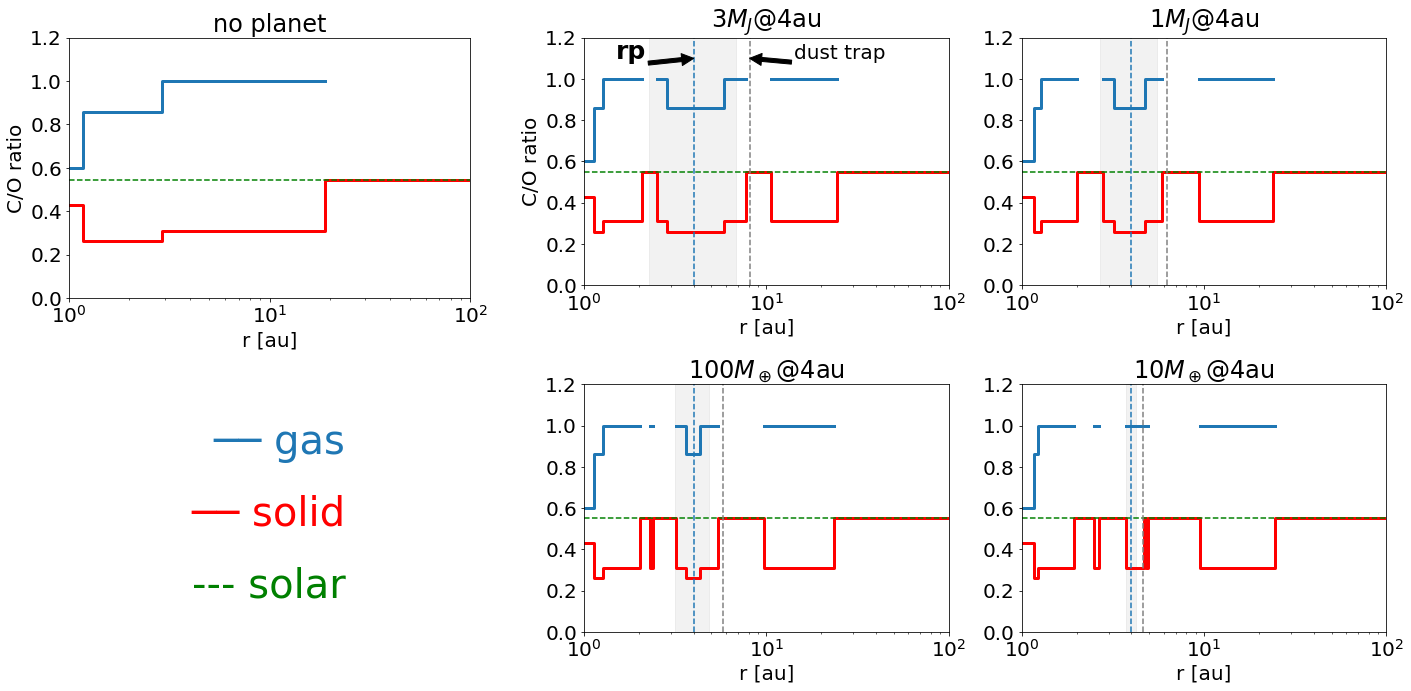} 
\centering
\caption{C/O ratio as a function of location in a disk, for different masses of a gap-opening planet at $4\,$au. From the upper middle to lower right panels, we show a planet mass of $3\Mj$, $1\Mj$, $100\Me$, and $10\Me$ at $4\,$au using the $\ite$ method, and a disk without planets (upper left panel) for comparison. The blue and red solid lines show gas- and solid-phase C/O ratios, whereas the solar C/O ratio is marked by a green dashed line.
The gap regions are shaded. The planet location and dust trap location are marked as vertical cyan and grey lines. 
}
\label{fig:CtoO} 
\end{figure*}

To investigate how radial variations in the C/O ratio are affected by feedback from gap-opening planets, and the presence of a puffed-up inner rim, we follow the prescription from \citet{oberg_effects_2011}. We assume the only C and O carriers are H$_2$O, CO, CO$_2$, refractory carbon, and silicate minerals, using the same abundances as that study. The total abundance of each species summed over the gas and solid phase does not vary with radius.

Figure \ref{fig:CtoO} shows how the gas- and solid-phase C/O ratio varies as a function of location in a disk, and for different $\Mp$ at $\rp = 4\,$au, for models using our iteration method. For comparison, the baseline model without a planet is also shown, analogous to the standard ``\"{O}berg model'' for the C/O profile.  We note that analogous changes can be observed for planets at larger orbits, but due to the relevance to most known planetary systems which are close-in, as well as the analogy with the solar system, we focus here on the $4\,$au case.

Based on our iterative models, the introduction of a gap-opening planet 
significantly alters the radial profile of the C/O ratio in the gas and solid phase, compared to the baseline (no-planet) case. 

Firstly, the presence of a gap makes the disk temperature (and pressure) profile strongly non-monotonous, which can create multiple iceline locations for a single chemical species. A monotonous, smoothly decreasing temperature profile underlies the widely studied picture of well-defined, unique icelines. In that case, the more refractory species (silicates, organic carbon, water ice) each have their iceline closer to the star than the more volatile species (e.g., CO).

Secondly, by comparing the results from disks hosting different mass planets, we can see the planet gap-opening effect on C/O is stronger as planet mass increases. 
The reduced optical depth within the gap leads to increased heating which causes $\Tmid$ to rise above the CO$_2$ sublimation temperature. This returns proportionally more oxygen than carbon back to the gas phase, thereby decreasing the gas-phase C/O ratio locally.

Our results show that the feedback from gap-opening planets can significantly affect the gas- and solid-phase C/O ratio at small spatial scales within a protoplanetary disk. Such variations have important implications for the composition of icy planetesimals, and the gas from which giant planets accrete their envelopes. Additionally, our findings demonstrate that radially distinct regions of the disk can be characterised by the same C/O ratio, which complicates the usage of C/O as a formation tracer. To construct more accurate models, it is essential that future observations focus in measuring the C/O ratio at planet-forming scales. We note, however, that some of the variations seen in the radial location of molecular icelines in our models are as little as $\sim1$ au, which can be difficult to resolve even with ALMA, though the larger shifts ($\sim10\,$au or more) can be more easily measured. The largest-scale variations are evident for high-mass planets at large separations ($M_\text{P}=3M_\text{J}$ at 30 au in our model, Figure \ref{fig:ice}a).

The degree to which the gap-modified gas and solid composition will be reflected in the atmospheric composition of a forming planet will further depend on the degree of mixing between the core and atmosphere, and the amount of sublimation that takes place during accretion. The scenario is further complicated by considering the vertical layer in which planets accrete their envelopes. Meridional flows from the disk surface may favour the accretion of gas and small grains from the disk surface layers, for example \citep[e.g.][]{teague_meridional_2019}.

\subsection{Ice lines and dust rings}

The altered thermal and ice line structure of a disk with a gap-opening planet has implications for the observational study of disk substructure, both spectroscopy of the gas and also the dust rings which are widely observed in disks with ALMA. 

We illustrate this for the case outlined in Figure \ref{fig:T4au}(a), a 3$\Mj$ planet at 4 au. Heating due to gap-opening increases the local midplane temperature above the CO sublimation temperature, introducing a new CO condensation front at the outer edge of the gap in a region of the disk where CO would otherwise be entirely frozen out.

As also highlighted in Figure \ref{fig:T4au}, a dust trap is located in the pressure maximum just outside a gap. Furthermore, results in the literature suggest regions near ice lines may be favorable for the pile-up of icy pebbles \citep[e.g.,][]{hyodo_formation_2019}. As pebbles cross the ice line and sublimate, outward diffusion followed by recondensation may locally enhance the surface density outside of the iceline, triggering instabilities which can lead to rapid pebble and planetesimal growth \citep{drazkowska_planetesimal_2017}. High dust-to-gas ratios and viscosity gradients produced by the density enhancement could further amplify the effect \citep[e.g.,][]{brauer_planetesimal_2008, ros_ice_2013, bitsch_stellar_2014, drazkowska_can_2014, flock_gaps_2015}.

This rapid growth of pebbles around condensation fronts is tentatively supported by observations of disks such HL Tau, where the location of millimeter dust rings has been linked to the icelines of water and other key volatiles \citep{zhang_evidence_2015}. Similarly, grain size distributions inferred from ALMA observations of HD 163296 are consistent with the enhanced production of large grains at the CO iceline \citep{guidi_dust_2016}. However, no unambiguous correlation between dust rings and ice lines on a standard monotonously radially decreasing temperature profile has been found. 
Results using empirical temperature estimates seem to disfavor such correlation \citep{long_gaps_2018}.

Although icelines have been invoked to explain the rings and gaps observed in a handful of disks, such as HL\,Tau \citep{zhang_evidence_2015}, icelines are not a preferred explanation when looking at large surveys of protoplanetary disks \citep{huang_disk_2018, long_gaps_2018, van_der_marel_protoplanetary_2019}. This is because most of the locations of substructures do not coincide with the sublimation temperature of the main disk volatiles, when assuming that the disk temperature is set by stellar irradiation. Under this hypothesis, a correlation between the location of substructures and the stellar luminosity is expected. 

However, as we show in this work, this potential correlation may get much more complicated when a planet is embedded in the disk. The planet alters the temperature-pressure profile of the disk, moving the ice lines to different radii and even creating multiple, radially widely separated ice lines for a single species. 
Therefore, our current results suggest that it is not necessarily a correlation with the stellar luminosity as it is usually assumed, but that embedded gap-opening planets need to be accounted for to fully assess the locations of ice lines and their correlation with the locations of dust (pebble) rings.

In addition, \citet{pinilla_dust_2017} demonstrated that due to the variations of dust sticking properties, ice-covered dust particles can create ``traffic jams'', which result in rings and gaps when observed at different wavelengths. The inclusion of dust evolution models in the framework of our models is needed to test if multiple substructures are expected in the disks as a result of a single planet embedded and multiple icelines locations of different volatiles. 

\subsection{Inner rim midplane temperature drops} \label{sec: T drop}

In this section, we discuss the temperature decrease in the inner few au in disks in section \ref{sec: Tmid}. As Figure \ref{fig:T4au}, \ref{fig:T10au}, and \ref{fig:T30au} show, even the lowest $\Mp= 10\Me$ case which represents minor or no planet effect on the disk, the iteration method predicts strong $\Tmid$ drops in this inner disk region. On the contrary, such an effect is not shown in the non-iteration method. The physical explanation is that the puff-up of the scale height can cause a strong shadowing effect to decrease the temperature within 10 au in the $\radmc$ simulation.
The reason why there is a puffed-up scale height at the inner rim at 1 au in our transition disk models is that stellar photons hit a dust wall, increasing the dust temperature.

As we assume the dust and gas temperatures are well coupled, the gas temperature is also high and causing the gas scale height to have a strong puff-up.
This phenomenon is also suggested in Figure 3 in \citet{dullemond_passive_2001}, as well as \citet{jang-condell_gaps_2012, jang-condell_gaps_2013, siebenmorgen_shadows_2012, zhang_self-consistent_2021}.
In our $\radmc$ setups, we input time evolving scale height for the surface density to volume density extension process. As the scale height indicated by the $\Tmid$ from last $\radmc$ has puff-up in the inner rim, our iteration models can naturally capture such effects. However, in the non-iteration method, the input scale height for $\radmc$ is just the smooth flaring scale height as that in the $\fargo$ setup. Therefore, the $\ite$ can have an advantage in making use of the physical temperature obtained by $\rt$ for a specific disk model rather than using the initially assumed temperature as $\noit$.
In addition, we measure the aspect ratio $\hpr$ of the puff-up inner rim at 1 au is about 0.035 and then it decreases to the lowest value of about 0.015 at about 1.5 au. 
For the region further away from 1.5au, $\hpr$ increases as a power law with a flaring index of 0.25 which is similar to the power law profile of $\noit$ $\hpr$.

The change in inner disk temperature structure between the commonly used isothermal (non-iteration) method and our iteration method also impacts the behaviour of elemental ratios like C/O. In our models using the iteration method, shadowing by the puffed-up inner rim causes $\Tmid$ dropping off more quickly within the inner disk, moving the H$_2$O and CO$_2$ icelines inwards. This translates to a steep rise in the gas-phase C/O ratio, as a large proportion of the total atomic oxygen is frozen-out into solids. In this scenario, C/O reaches unity within $\sim 1.5$ au, compared to $\sim 3$ au in the $\noit$ and classical models.

\subsection{Assumptions and limitations} 

There are a number of simplifications in our $\hd$ and $\rt$ simulations that can be improved in future work. 
First, we only consider 2D $\hd$ simulations in radial and azimuthal directions instead of full 3D $\hd$ simulations, which benefits us for speeding up the whole $\ite$ process. However, 3D $\hd$ simulations can allow one to get rid of the vertical isothermal assumption which will be useful for addressing vertical stratified problems (e.g. ice-surface distribution, gas molecule emission layers.) In this paper, we only focus our discussion on the midplane temperature and its effect on midplane ice distribution. 

Second, we have some simplifications about dust in our modelings. In $\hd$ simulations, we do not include dust species in order to speed up the simulation process. 
In addition, only one small grain size, 0.1$\mu m$, is included in the $\rt$ simulations. 
If we consider grain size distribution or dust evolution process, like grain growth or fragmentation, it is still unclear how can these factors change the dust distribution and hence disk temperature.
Because we do not have a dust density distribution in our models, we also neglect dust settling in our models. As a consequence, it is possible that when dust settling is included, less dust remains on the disk surface, allowing stellar radiation to penetrate deeper into the disk and increase $\Tmid$. 
Our models also neglect to account for dynamical effects such as radial drift and mass accretion, which add considerable complexity. For example, studies have shown that radial drift can produce multiple icelines \citet{cleeves_multiple_2016} or make icelines thermally unstable under typical disk conditions \citep{owen_snow_2020}. Icy volatiles drift faster that those in the gas-phase, which results in the iceline progressively moving inwards, condensing more volatiles. The iceline then recedes as volatiles sublimate, on timescales much shorter that the disk lifetime (1000-10,000 years). Similarly, the mass accretion rate plays an important role in iceline evolution, with iceline moving inwards when accretion rates are high, and migrating back out in the later stage of disk evolution when the accretion rate decreases \citep{oka_evolution_2011}. The combined effects of radial drift and mass accretion can cause molecular icelines to moves inwards by as much as 60\% \citep{piso_co_2015} 

Third, we do not consider viscous heating which can be dominant in the midplane of the inner disk (e.g. \citet{broome_iceline_2022}). Thus, the viscous heating may have sufficient effects on increasing the very inner disk $\Tmid$. This may have a strong effect on our 4au cases.  Also, our models do not capture shock heating from the planet which can be significant for massive planet cases.

Finally, we choose 100 orbits as our iteration step to implement the feedback from $\radmc$ to $\fargo$. 
However, we notice that the thermal relaxation time can vary from about 100 to 0.1 dynamic timescale from 1 to a few tens au \citep{malygin_efficiency_2017, pfeil_mapping_2019} and our model can not capture this. 
The number of the iteration step we decide is a balance between the total simulation time and reflecting the gap opening thermal feedback properly. 
One possible way to improve the approach is if we are only concerned about the radial temperature structure but ignore the azimuthal variations, we can use fewer photon package numbers for fewer azimuthal grid cell $\radmc$ simulations to speed up each iteration step and do more iterations. 

In future work, we will focus on the improvement of some of these limitations, in particular the effect of including dust in the models.

\section{Conclusions}\label{sec:conclusion}
In this paper, we present a new method to study the gap-opening effect on protoplanetary disk temperature structure by iterating $\hd$ and $\rt$ simulations.  We quantify the planet-opening gap profiles including gap width,  depth, and eccentricity, and explore the dust-trapping condition in outer gap edges. By obtaining the temperature profiles in disks, we study the volatile iceline locations and ultimately provide new C/O ratio  for disks with embedded planets. During the modeling, we compare our $\ite$ models with the conventional $\noit$ models and conduct parameter studies of different planet masses $\Mp$ and planet locations $\rp$. Our main conclusions are as follows: 

(i) Gap profiles: the $\ite$ method predicts deeper and more eccentric gaps than the $\noit$. The most significant difference in gap depth comparison between these two methods is seen at 1$\Mj$ at 4 au or 10 au, where the $\ite$ gap depth is about an order of magnitude deeper than the $\noit$. 

(ii) Dust trap: both $\ite$ and $\noit$ indicate similar locations of pressure maximum for dust trapping $\rdt$, which is about 7-10 $\rhill$ further away from $\rp$. 
However, the iteration predicts a larger fragmentation grain size across the pressure bump, and as a consequence, a slightly higher fraction of dust could be trapped in the pressure bump.

(iii) Midplane temperature: our $\ite$ models can capture the gap-opening process by a planet and its effect on the time evolution of the disk temperature structure, whereas the conventional $\noit$ models do not capture.
By implementing the $\ite$  method, we show that the strong midplane temperature drops in the inner few au of disks because of the shadowing effect caused by the puff-up disk inner rim.  Meanwhile, the midplane temperature $\Tmid$ increases significantly in the gap regions due to more photons can penetrate into the midplane. The maximum $\Tmid$ contrast between gap edges and gap center is about 40K when $3\Mj$ or $1\Mj$ presents at 4 au. 

(iv) Icelines: because of $\Tmid$ drops, \ce{CO2} and \ce{CO} ice may exist in the inner disk region (at a few au) in the $\ite$ model.  At gap regions, both iteration and $\noit$ predict that \ce{CO2} or \ce{CO} ice may  sublimate. As a result, the $\noit$ suggests that more than one \ce{CO2} or \ce{CO} iceline can appear in a disk, whereas the $\ite$ models propose that even more \ce{CO} icelines might exist in giant planet-forming disks. This result suggests that the combination of an embedded planet and different locations of the iceline of the same volatile can still explain the observed substructures in protoplanetary disks. 

(v) C/O ratio (Fig.~\ref{fig:CtoO}): the radial C/O ratio profile across the disk is significantly more complex due to the presence of gaps opened by giant planets in comparison to disk models without embedded planets. As a consequence, the feedback of the planet-opening gap on the disk thermal structure can significantly influence the composition of material available for the giant planet's atmosphere or for the next generation of planet formation.

\section*{Acknowledgments}
We are thankful to the referee for the constructive report. We thank Jaehan Bae, Jeffrey Fung, Min-Kai Lin, Alexandros Ziampras for help and useful discussions. K.C. acknowledges support by UCL Dean's Prize and China Scholarship Council. L.K. acknowledges funding via a Science and Technology Facilities Council (STFC) studentship. M.K. has received funding from the European Union's Horizon Europe research and innovation programme under grant agreement No. 101079231 (EXOHOST), and from UK Research and Innovation (UKRI) under the UK government’s Horizon Europe funding guarantee (grant number 10051045).

%%%%%%%%%%%%%%%%%%%%%%%%%%%%%%%%%%%%%%%%%%%%%%%%%%

\section*{Data Availability}
Data from our numerical models are available on reasonable request to the corresponding author.
The $\fargo$ code is publicly available from
\href{https://fargo3d.bitbucket.io/download.html}{https://fargo3d.bitbucket.io/download.html}.
The $\radmc$ code is available from \href{https://www.ita.uni-heidelberg.de/~dullemond/software/radmc-3d/}{https://www.ita.uni-heidelberg.de/~dullemond/software/radmc-3d/}.

%%%%%%%%%%%%%%%%%%%% REFERENCES %%%%%%%%%%%%%%%%%%

% The best way to enter references is to use BibTeX:

\bibliographystyle{mnras}
% \bibliography{zotero1.bib} 
% \bibliography{thispaper.bib} 

%%%%%%%%%%%%%%%%%%%%%%%%%%%%%%%%%%%%%%%%%%%%%%%%%%

%%%%%%%%%%%%%%%%% APPENDICES %%%%%%%%%%%%%%%%%%%%%

\appendix

\section{Tests of iteration steps}
\label{app:iteration steps}

\begin{figure*}
\includegraphics[width=\linewidth]{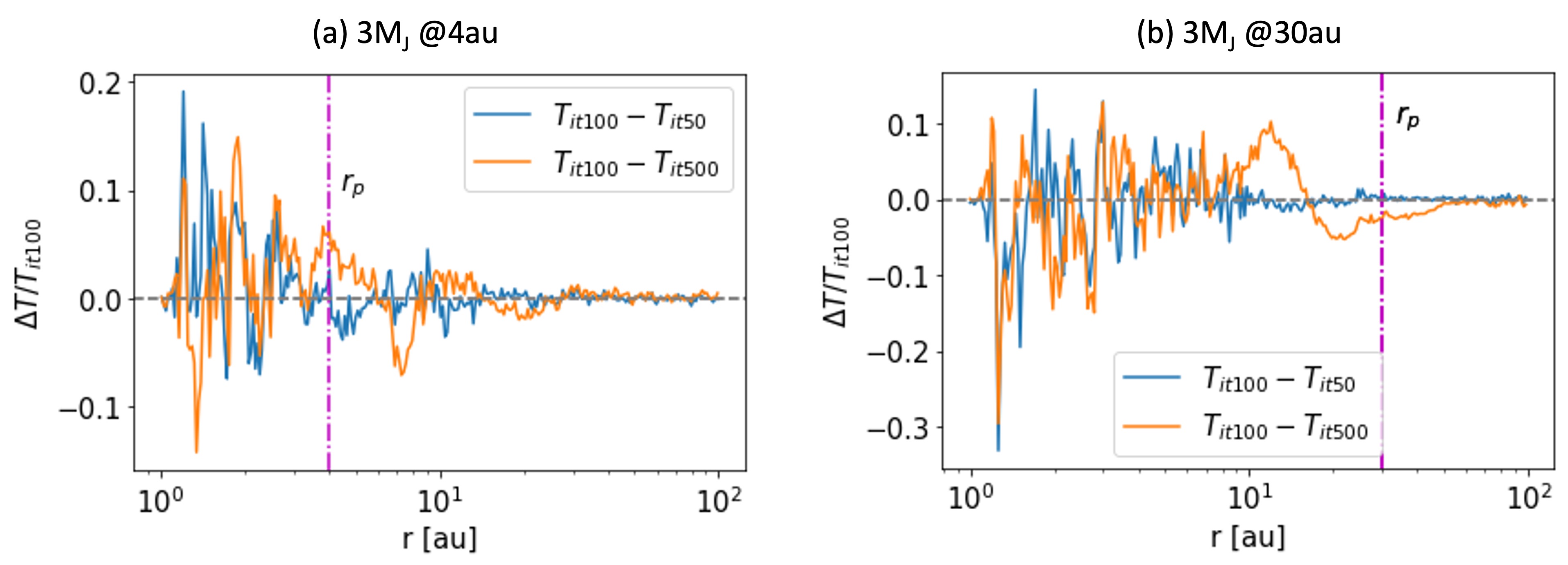} 
\caption{Comparison of midplane temperature over 2000 orbits by using different iteration steps. The y-axes $\Tratio$ show the normalized temperature difference between two iteration steps. $T_{\rm{it50}}$, $T_{\rm{it100}}$, or $T_{\rm{it500}}$ is the temperature by adopting iteration steps of 50, 100, or 500 orbits, respectively.
}
\label{fig:it_test} 
\end{figure*}

We compare the midplane temperature after 2000 planetary orbits by implementing iteration steps of 50 orbits, 100 orbits, or 500 orbits in Figure \ref{fig:it_test}. We show two case of 3$\Mj$ at 4 au (left) and 3$\Mj$ at 30 au (right). The differences between the iteration step of 100 orbits and 50 orbits displayed by blue lines are small, especially at the gap regions, the normalized temperature difference $\Tratio \sim 0$ and the maximum is $<0.1$.
On the other hand, the differences between the iteration step of 100 orbits and 500 orbits show relatively larger fluctuations around 0. Therefore, the iteration step of 100 orbits basically is able to reproduce the iteration step of 50 orbits, whereas the iteration step of 500 orbits could not reproduce the iteration step of 100 orbits very well. 
Meanwhile, we acknowledge that the thermal relaxation time can vary more than several magnitudes in different disk radii \citep{malygin_efficiency_2017, pfeil_mapping_2019} but our model can not capture this. Nevertheless, due to the limit of computing capability, we think adopting 100 orbit as the iteration step is suitable for our studies. 

\section{Comparisons between midplane temperature and density-weighted vertical averaged temperature}
\label{app:Tmid_vs_Tavgz}

\begin{figure*}
\includegraphics[width=0.5\linewidth]{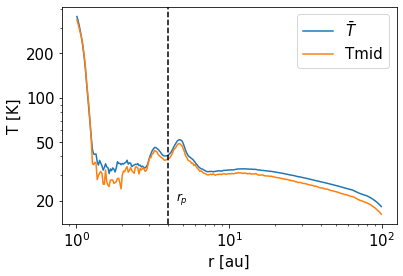} 
\caption{Comparison of midplane temperature and the density-weighted vertical averaged temperature from $\radmc$ simulations of 3$\Mj$ at 4au over 100 orbits.
}
\label{fig:Tavz} 
\end{figure*}

We compare the midplane temperature and the density-weighted vertical averaged temperature from $\radmc$ simulations of 3$\Mj$ at 4au over 100 orbits in Figure \ref{fig:Tavz}. The density-weighted vertical averaged temperature is calculated by 

\begin{equation}
\bar{T}(r)=\frac{\int \rho(r, z) T(r, z) d z}{\int \rho(r, z) d z}
\end{equation}
where $T(r, z)$ is the azimuthal average temperature. 

Overall, the density-weighted temperature is not significantly different from the midplane temperature as the volume density is much higher in the midplane than on the surface. Especially the gap region temperature is very similar.  
The strongest difference is in the shadowing region where the midplane temperature is lower than the weighted temperature for up to 10K. This is because the shadowing effect is strongest for the midplane. As the weighted temperature has a contribution from the surface temperature, which is not heavily affected by the shadowing effect, it makes the weighted temperature higher.  

\section{Gas density of planets at 10au and 30au} 

%% density rp=10au
\begin{figure*}
\includegraphics[width=\linewidth]{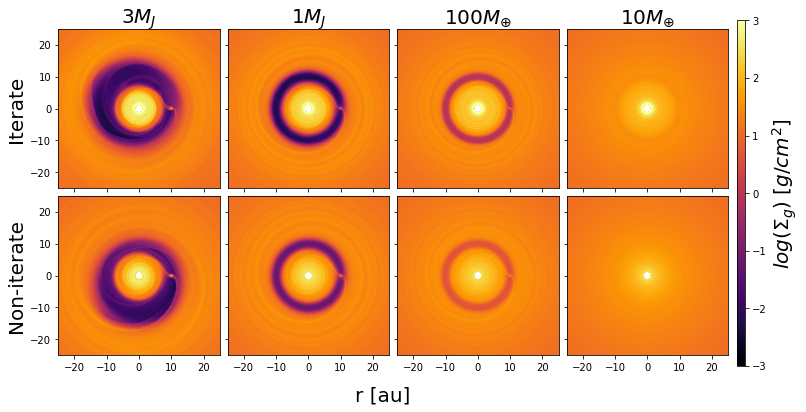} 
\caption{Similar to Figure \ref{fig:dens4au} bur for planets at 10 au.
}
\label{fig:dens10au} 
\end{figure*}

\begin{figure*}
\includegraphics[width=\linewidth]{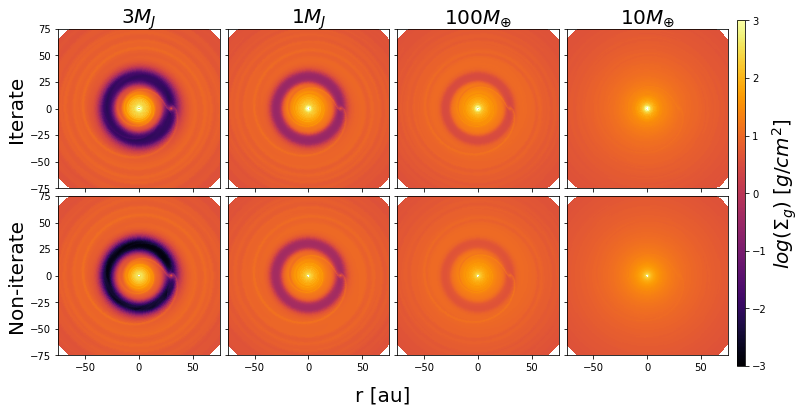} 
\caption{Similar to Figure \ref{fig:dens4au} bur for planets at 30 au.
}
\label{fig:dens30au} 
\end{figure*}

Figure \ref{fig:dens10au} and \ref{fig:dens30au} show the 2D gas density map of gaps opened by 3$\Mj$, 1$\Mj$, 100$\Me$, and 10$\Me$ planets at 10 au and 30 au, respectively. The results of $\ite$ and $\noit$ are displayed in the upper and lower panels, respectively.

\section{Temperature of planets at 10au and 30au} 

\begin{figure*}
\includegraphics[width=\linewidth]{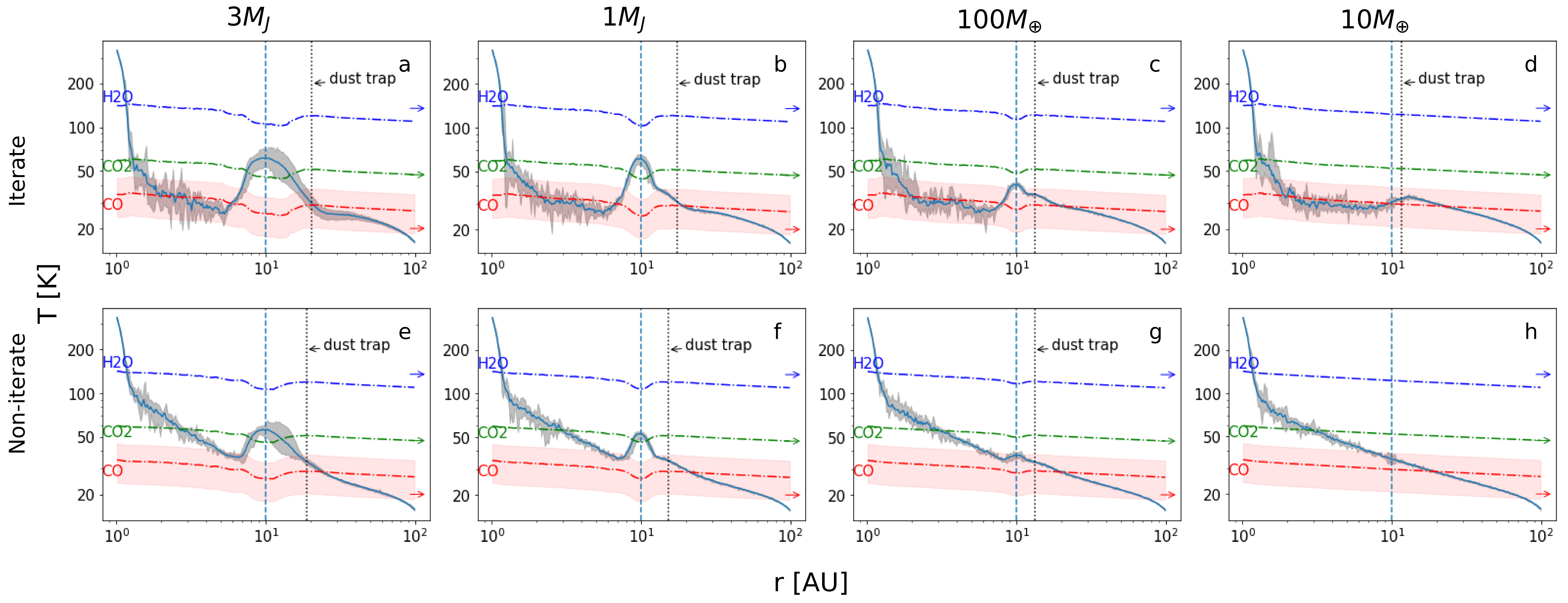} 
\caption{Similar to Figure \ref{fig:T4au} but for planets at 10 au. 
}
\label{fig:T10au} 
\end{figure*}

\begin{figure*}
\includegraphics[width=\linewidth]{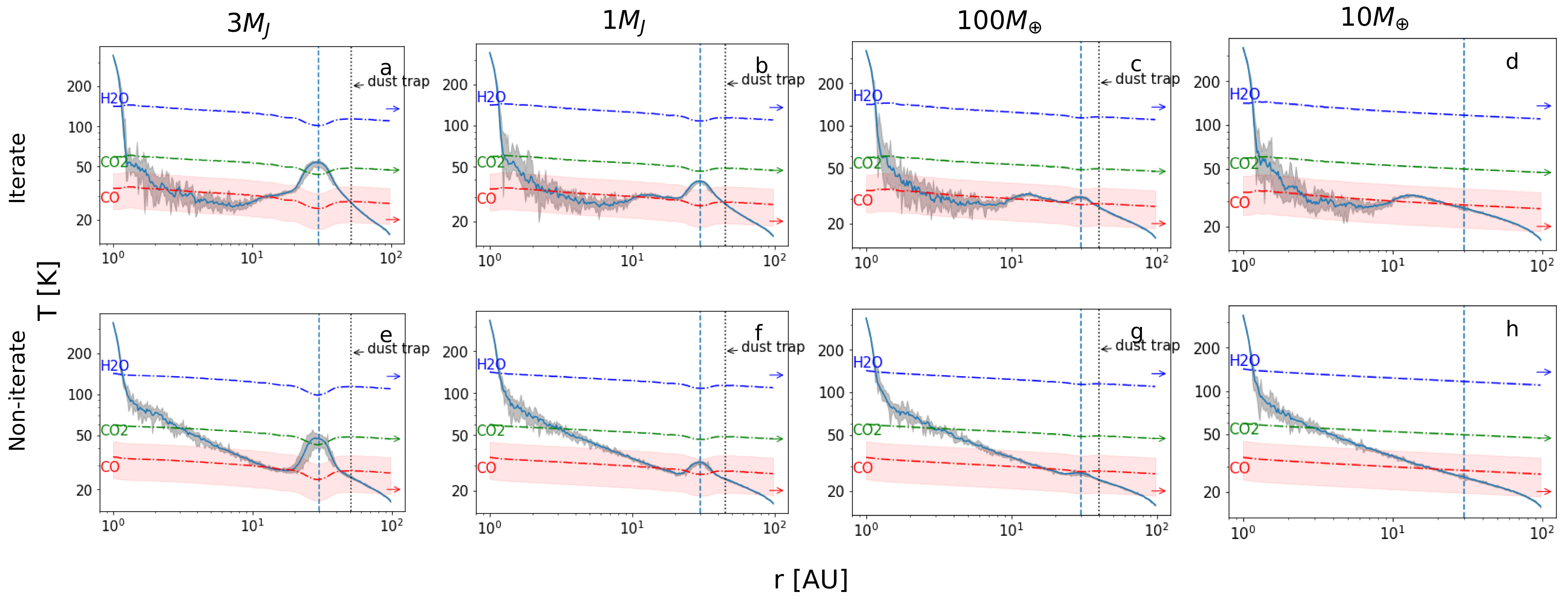} 
\caption{Similar to Figure \ref{fig:T4au} but for planets at 30 au. 
}
\label{fig:T30au} 
\end{figure*}

Figure \ref{fig:T10au} and \ref{fig:T30au} show the midplane temperature as a function of disk radius when planet location is $\rp$ = 10 and 30 au, respectively. 

%%%%%%%%%%%%%%%%%%%%%%%%%%%%%%%%%%%%%%%%%%%%%%%%%%

% Don't change these lines
\bsp	% typesetting comment
\label{lastpage}
\end{document}